\documentclass[useAMS,usenatbib]{mn2e}
\usepackage{graphicx}
\usepackage{natbib}
\usepackage{epstopdf}
\usepackage{amsbsy}
\usepackage{amssymb}
\usepackage{pifont}
\usepackage{amsfonts}
\usepackage{color}

\newcommand{\lya}{Ly-$\alpha$~}

\providecommand{\adsurl}[1]{\href{#1}{ADS}}

\def\aap{A\&A}
\def\apj{ApJ}

\def\apjl{ApJ}
\def\mnras{MNRAS}

\def\nat{Nat}

\def\apjs{ApJS}

\def\HI{\hbox{H$\,\rm \scriptstyle I\ $}}

\def\HeII{\hbox{He$\,\rm \scriptstyle II\ $}}

\DeclareMathAlphabet{\mathsc}{OT1}{cmr}{m}{sc}
\def\testbx{bx}%
\DeclareRobustCommand{\ion}[2]{%
\relax\ifmmode
\ifx\testbx\f@series
{\mathbf{#1\,\mathsc{#2}}}\else
{\mathrm{#1\,\mathsc{#2}}}\fi
\else\textup{#1\,{\mdseries\textsc{#2}}}%
\fi}

\title[Line broadening in the \lya forest]{Diagnosing galactic
  feedback with line broadening in the low redshift Lyman-$\alpha$
  forest} \author[M. Viel et al.]  {Matteo Viel$^{1,2,3}${\thanks{email:
      viel@oats.inaf.it}}, Martin G. Haehnelt$^4$ , James
  S. Bolton$^5$, Tae-Sun Kim$^1$, \newauthor  Ewald Puchwein$^4$, Fahad
  Nasir$^5$ \& Bart P. Wakker$^6$ \\ $^1$ INAF - Osservatorio
  Astronomico di Trieste, Via G.B. Tiepolo 11, I-34131 Trieste, Italy \\
  $^2$ SISSA, Via Bonomea 265, 34136 Trieste, Italy\\ $^3$ INFN/National Institute for Nuclear Physics, Via Valerio 2,
  I-34127 Trieste, Italy\\ $^4$ Kavli Institute for Cosmology and
  Institute of Astronomy, Madingley Road, Cambridge, CB3 0HA,
  UK\\ $^5$ School of Physics and Astronomy, University of Nottingham,
  University Park, Nottingham, NG7 2RD, UK\\ $^6$ Supported by NASA/NSF, affiliated with Dept. of
  Astronomy, University of Wisconsin-Madison, 475 N. Charter Street,
  Madison, WI 53706, USA}

\begin{document}
\maketitle
 
\begin{abstract}
We compare the low redshift ($z\simeq 0.1$) \lya forest from
hydrodynamical simulations with data from the Cosmic Origin
Spectrograph (COS). We find tension between the observed number of
lines with $b$-parameters in the range $25$--$45$ $\rm km\,s^{-1}$ and
the predictions from simulations that incorporate either vigorous
feedback from active galactic nuclei or that exclude feedback
altogether.  The gas in these simulations is, respectively, either too
hot to contribute to the \lya absorption or too cold to produce the
required line widths.  Matching the observed $b-$parameter
distribution therefore requires feedback processes that thermally or
turbulently broaden the absorption features without collisionally
(over-)ionising hydrogen.  This suggests the \lya forest $b$-parameter
distribution is a valulable diagnostic of galactic feedback in the low
redshift Universe.  We furthermore confirm the low redshift \lya
forest column density distribution is better reproduced by an
ultraviolet background with an \HI photo-ionisation rate a factor
$1.5$--$3$ higher than predicted by \citet{hm12}.

\end{abstract}


\section{Introduction}
\lya forest data have become an important tool in studying the
physical state of the intermediate redshift ($2<z<5$) intergalactic
(IGM) and circumgalactic medium. With the advent of the Cosmic Origins
Spectrograph (COS) on the \emph{Hubble Space Telescope (HST)}, it has
become possible to obtain much improved measurements also at lower
redshift \citep{savage14,shull14,danforth16,werk16,pachat}. The
increased resolution and signal-to-noise (S/N) of the COS data enable
the measurement of the column density of \lya absorbers to lower
values and help resolve the thermal broadening for weaker absorbers,
complementing earlier investigations of the low-redshift IGM
\citep{weymann98,janknecht06,kirkman07}.

Concurrently, the interpretation of these data has been aided by high
dynamic range cosmological hydrodynamical simulations incorporating
much of the relevant (sub-grid) physics at $z=0$
\citep{torna10,dave10,tepper,ford13,villa16,rahmati16}.  The present
consensus on the nature of these absorbers is that they trace galactic
environments relatively faithfully and may be used to address a wide
set of scientific questions, from finding the missing baryons to the
nature of the ultraviolet background (UVB) and galactic feedback.
Here, we compare a new measurement of the observed \HI \lya Doppler
$b$-parameter and column density distribution at $z=0.1$ with
predictions from a range of state-of-the-art numerical simulations.
We assess whether constraints on the physical mechanism responsible
for stellar and active galactic nuclei (AGN) feedback may be obtained,
and revisit the possible missing ionising photon problem first
discussed by \citet{kollmeier14} and further investigated by
\citet{shull15}, \citet{wakker15}, \citet{khaire} and \citet{bird16}.

\section{COS data}
We have selected 44 {\it HST}/COS AGN spectra available as of December
2015 in the {\it HST} MAST (Mikulski Archive for Space
Telescopes). The two main selection criteria are: a S/N per resolution
element that is larger than 20 and an emission redshift in the
redshift range $0.1 < z < 0.35$, covering the \lya forest at $0 < z
< 0.2$.  The first criterion was imposed so that the detection limit
is $\log N_{\mathrm{\ion{H}{i}}}$/cm$^{-2} \sim 13$.  The final
co-added COS spectra have a resolution of $\sim 18$--20 km s$^{-1}$ in
a heliocentric velocity frame and have $\rm S/N \in [30-150]$ per
resolution element in the \lya forest region. The total redshift
coverage is $\Delta z=4.991$ excluding Milky Way interstellar medium
line contamination and unobserved wavelength regions.  Details of the
COS data reduction and the properties of the AGN spectra can be found
in \citet{wakker15} and Kim et al. (2016, {\it in prep.}),
respectively.

After initial continuum fitting, all the absorption profiles were
identified and fitted with a Voigt profile using {\tt VPFIT}
\citep{carswell14} to obtain the column density and the $b-$parameter
\citep[see][for more details]{kim13,kim16}. {\tt VPFIT} is also used
to obtain line parameters for our simulated spectra.  Since the
simulated spectra are fitted only with \ion{H}{i} \lya lines, we have
also fitted the observed \lya lines without using any higher-order
Lyman series lines. Depending on the date of the observation, a
non-Gaussian COS line spread function (LSF) at the different lifetime position was used \citep{kriss11}.
At $0 < z < 0.2$, the total number of fitted \ion{H}{i} lines is 704
at $\log N_{\mathrm{\ion{H}{i}}}$/cm$^{-2} \in [12.5, 14.5]$, with the
$b$-parameters spanning the range 8--181 $\rm km \,s^{-1}$. There are
424 lines with $\log N_{\mathrm{\ion{H}{i}}}$/cm$^{-2} \in [13, 14]$ 
with a relative error on the $b$-parameter smaller than 0.5: this will
constitute our main sample. For comparison, we shall also use the \lya
lines obtained by \cite{danforth16} from 39 COS AGN ($\Delta z =
4.33$).  We find good agreement between the data set used here and
the one presented in \cite{danforth16} (D16), as will be demonstrated
later.

\section{Numerical simulations}
We consider a range of state-of-the art $\Lambda$CDM cosmological
hydrodynamic simulations including the Illustris \citep{illustris,nelson} and
Sherwood \citep{bolton17} simulations.  The majority of the
simulations have been performed with the parallel Tree-PM smoothed
particle hydrodynamics (SPH) code {\sc P-Gadget-3} \citep{springel},
apart from Illustris that was run with the moving-mesh code {\sc AREPO}
\citep{springel10}. The simulations include a variety of star
formation and stellar or AGN feedback implementations as well as a
range of UVB models.   We have also boosted the \HeII photo-heating rates in
some models in an ad hoc manner \citep[as described in][]{bolton08pdf}
to obtain temperatures for the low density, photo-ionised IGM that
better match the observed $b$-parameter distribution.  The main
properties of the individual simulations are as follows.

{\bf HM (Haardt \& Madau UVB models)}.  These are {\sc P-Gadget-3}
simulations with a range of assumptions for the UVB and
temperature of the low density IGM. HM simulations are performed
without feedback using a simplified star formation criterion that
turns all gas particles with a density above $\rho/\langle \rho
\rangle=10^{3}$ and a temperature below $10^5 \rm\, K$ into star
particles.  This feature is labelled {\tt QUICKLYA} and was first used
by \cite{vhs}.  The HM01 runs \citep{hm01} differ
from the HM12 simulations \citep{hm12} in the choice of precomputed
UVB model and hence \HI photo-ionisation rate, which is  $\Gamma$ /$(10^{-12}) = 0.035$ and 0.127 for HM01 and HM12, respectively (see Table 1).  In addition,
the thermal history for each simulation is labelled as ``hot'' or
``vhot'', indicating a different assumption for the gas temperature,
$T_0$, at the mean background density, which is in the range  $\log (T_0/K) = 3.7-4.1$.  All the HM models are run with
a linear box size of $60h^{-1}$ comoving Mpc and $2\times512^3$ gas
and dark matter particles. 

{\bf{Illustris.}} The Illustris simulation has a linear box size of
$75h^{-1}$ comoving Mpc and follows the evolution of $2\times1820^3$
gas cells and dark matter particles. The star formation and feedback
model uses supernovae-driven winds which scale with the velocity
dispersion of the host halo \citep{vogelsberger2013}. AGN feedback is
based on \citet{sijacki07} and uses two models -- radiatively
efficient and ``radio-mode'' -- depending on the black hole accretion
rate. In the latter case $7$ per cent of the accreted rest mass energy
is thermally injected into AGN bubbles. The individual injection
events are highly energetic, corresponding to roughly $0.01 M_{\rm BH}
c^{2}$.  Photo-ionisation and heating are followed using the
\citet{faucher09} UVB, and self-shielding and ionising flux from
nearby AGN are accounted for. This results in $\Gamma$ /$(10^{-12}) = 0.048$, $\log (T_0/K) = 3.7$ 
and relatively high temperature for gas at moderate overdensities $\log (T_0/K) = 6.2$  (see Table 1).

{\bf{Sherwood.}} The Sherwood simulation that we primarily use here was
performed with a linear box size of $80h^{-1}$ comoving Mpc
and $2\times 512^3$ particles. It employs the
star formation and feedback model described in \citet{puchwein13}.
This follows the star formation prescription of \citet{springel2003}
with a Chabrier initial mass function and supernovae driven winds with
velocities that scale with the escape velocity of the galaxy. The AGN
feedback is again based on \citet{sijacki07}, but with more modest
assumptions about the available energy; $2$ per cent of the accreted
rest mass energy is injected in the radio mode and individual events
are much less energetic, with $\gtrsim 2\times10^{-6} M_{\rm BH} c^2$.
In addition,  two further Sherwood runs at different resolution are used
for convergence testing (not shown in any of the figures). These use
the simpler {\tt QUICKLYA} treatment, and have the same box size of
$80h^{-1}$ comoving Mpc and have $2\times 512^3$ or $2\times 1024^3$
particles, respectively. This run has $\Gamma$/$(10^{-12}) = 0.035$, $\log (T_0/K) = 3.9$.

The cosmological parameters for all the simulations are in agreement
with either \cite{wmap9yr} or \cite{Planck2014}.  Simulated spectra
are extracted from all models at $z=0.1$ along 1000 random lines of
sight (our results have converged for this number of
spectra). Resolution effects are taken into account by convolution
with the COS LSF.  The S/N per resolution element is chosen to be 30.
The simulated spectra are then analysed with {\tt VPFIT} adapted for
de-convolution of the COS LSF, using the same procedure used to fit
the observational data.  Although we will show data for a wider range
of column densities, it is only the range between
$N_{\mathrm{\ion{H}{i}}}=10^{13-14}$ cm$^{-2}$ that we found to be
robust with regard to resolution and noise issues (we discuss this
further below).  Unless otherwise stated, we therefore scale the mean
transmitted flux of the spectra to match the observed column density
distribution function (CDDF) in this range.  This rescaling is
performed by modifying the optical depth in each pixel of the
simulated spectra by a constant, $A_{\rm f}$, such that
$\overline{F}_{\rm f}=\langle e^{-A_{\rm f}\tau_{\rm i}} \rangle $.
Table~\ref{tabsims} summarises the simulations along with some
quantities discussed in the following sections.

We have also performed a series of convergence checks on the
simulations.  With regard to mass resolution, when comparing the CDDFs
of the {\tt QUICKLYA} Sherwood runs (not shown in Table~\ref{tabsims})
we found agreement at the 15 per cent level in the range $\log(N_{\rm
  HI}$/cm$^{-2})=12.5-14.5$ while the $b$-parameter distributions
agree to within 20 per cent at $>20\,\rm km\,s^{-1}$. Regarding box
size effects, we found that the HM12, Sherwood and Illustris
simulations are all in very good agreement; box sizes of $60h^{-1}$
comoving Mpc are large enough to effectively probe the range of column
densities considered here.  The same holds for the $b$-parameter
distribution.  In terms of the sub-grid physics, when we compare a
simulation with the effective star formation model of
\citet{springel2003} (not shown in the table) with the {\tt QUICKLYA}
HM runs we find that the CDDF and $b$-parameter distributions are in
good agreement: the CDDFs agree within 10 per cent in the range
$\log(N_{\mathrm{\ion{H}{i}}}$/cm$^{-2})=12.5-14.5$, while the
$b$-parameter distribution agrees within 25 per cent over the whole
range.  Since these errors are smaller than the statistical
uncertainties of the data, for our purposes {\tt QUICKLYA} does not
significantly impact on the column density range considered here when
compared to a more detailed star formation model.  Finally, the
$b-$parameter distribution from Illustris converges within 10 per
cent when using S/N values in the range 20-40 per resolution element
(the reference case is 30) at 17-70 $\rm km\,s^{-1}$, while the CDDFs
agree within 0.05 dex in the range
$\log(N_{\mathrm{\ion{H}{i}}}$/cm$^{-2})=13-14.5$.

\begin{table}

\setlength{\tabcolsep}{4.5pt}

\begin{tabular}{llllllll}
\hline
Model&$\Gamma$&$T_{0} $&$T_{+}$ &$\overline{F}$&$A_{\rm f}$& $\overline{F}_{\rm f}$& $\Gamma_{\rm f}$\\ 
\hline
\emph{HM01}&0.127& 3.72 & 4.93 & 0.985 &  1.252 & 0.982 & 0.101 \\
\emph{HM01$_{\mathrm hot}$}&0.127& 3.99 & 5.00 & 0.989 & 2.007 & 0.981 & 0.063\\
\emph{HM01$_{\mathrm vhot}$}&0.127& 4.08 & 5.00  &  0.990 & 2.426 & 0.980 & 0.052 \\
\emph{HM12}&0.035& 3.71& 4.90& 0.964 & 0.408 & 0.981 & 0.087\\
\emph{HM12$_{\mathrm hot}$}&0.035& 3.97 & 4.94 & 0.972 & 0.624 & 0.981 & 0.057\\

\emph{Illustris}&0.048&  3.73 & 6.19 & 0.976 & 0.982 & 0.977 & 0.049\\
\emph{Sherwood}&0.035& 3.91 & 5.12 & 0.965 & 0.496 & 0.979 & 0.071\\

\hline
\end{tabular}
\vspace{-0.1cm}
\caption{Hydrodynamical simulations used in this work.  The columns
  list: the simulation name; the \HI photoionisation rate, $\Gamma$,
  in units of $10^{-12}\rm\,s^{-1}$; $T_0$, the median temperature at
  the mean density (log. units, volume weighted) calculated for a
  random sampling of gas at $\log (1+\delta)=[-0.1$,$0.1]$ and excluding gas
  hotter than 10$^5\rm\,K$; $T_+$, the median temperature for
  overdensities $\delta=[4-40]$ (log. units); the simulated mean transmitted
  flux; the rescaling factor $A_{f\rm }$
  applied to match the CDDF in the range $\log (N_{\rm HI}$/cm$^{-2}$) =[13-14]; the mean flux obtained; the new $\Gamma_{\rm
    f}=\Gamma/A_{\rm f}$ value inferred from the rescaling.  Quantities are at $z=0.1$.   The observed mean flux is $\overline{F}=0.983$ from \citep{danforth16}.}
\label{tabsims}
\end{table}

\section{Results}

\begin{figure*}
\begin{center}
\includegraphics[width=7.595cm,height=7.75cm]{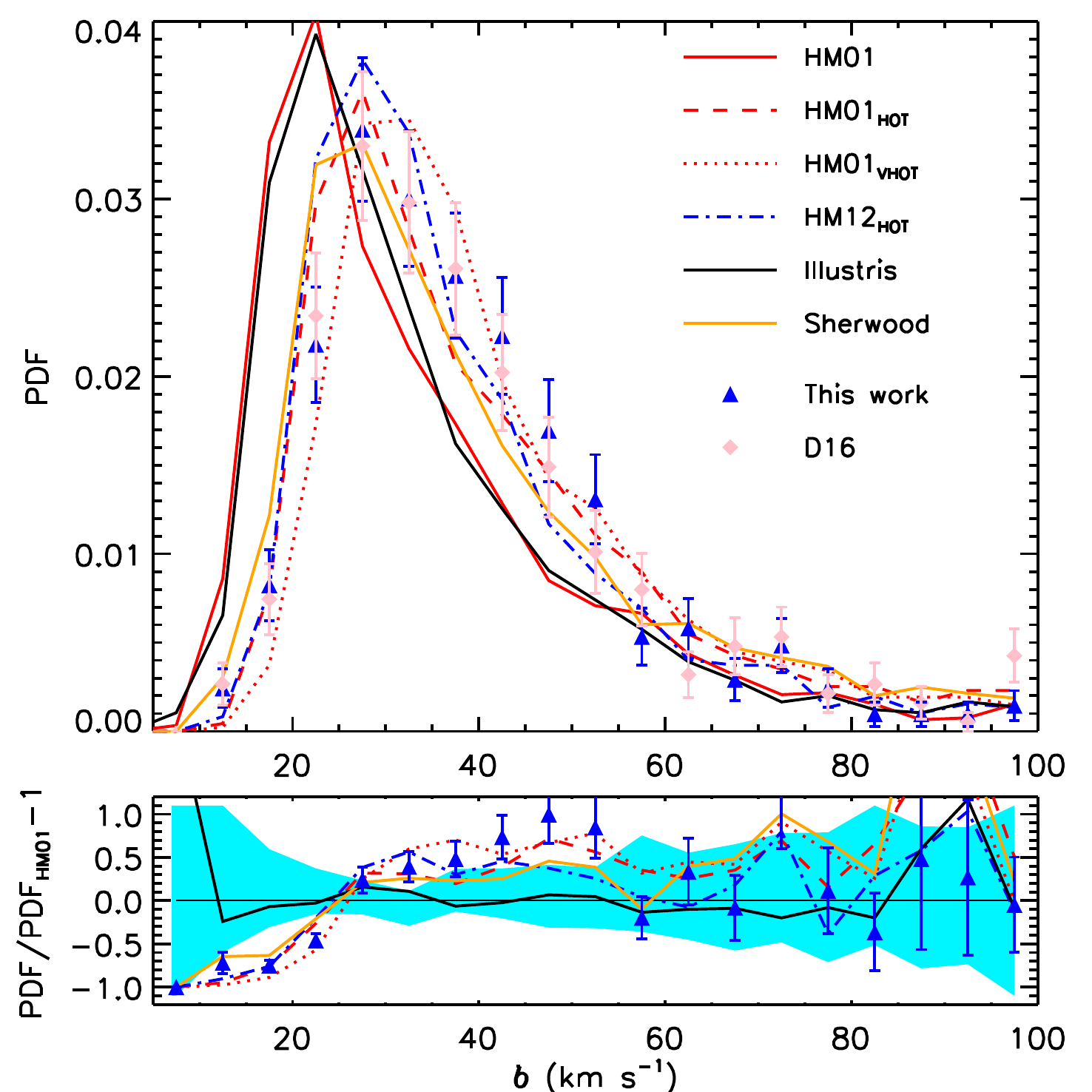}
\includegraphics[width=7.595cm,height=7.75cm]{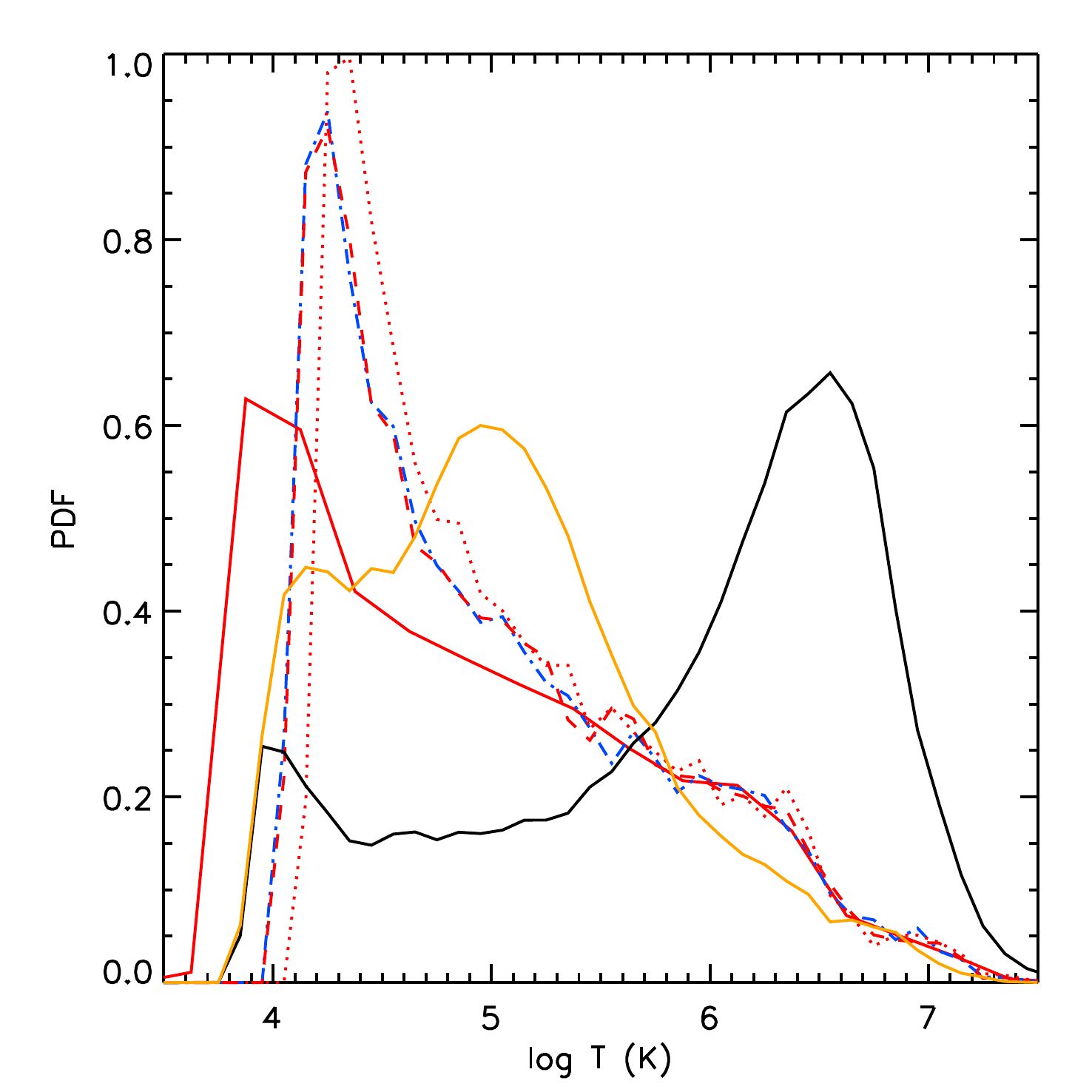}
\end{center}
\vspace{-0.4cm}
\caption{{\it Left.} The $b$-parameter distribution for HM01 (red
  solid line); HM01$_{\rm hot}$ (dashed red line); HM01$_{\rm vhot}$
  (dotted red line); HM12$_{\rm hot}$ (blue dot-dashed line);
  Illustris (black solid line); Sherwood (orange solid line). The
  bottom panel shows the ratio of the line width PDFs with respect to
  HM01.  The shaded area indicates the $\pm 2\sigma$ range obtained
  from a set of 100 mocks with the same redshift path as the data. COS
  data are represented by the blue triangles (Poisson error bars),
  while the D16 data are shown as pink diamonds.  The spectra have
  been scaled to match the observed CDDF at N$_{\rm HI}=10^{13-14}$
  cm$^{-2}$, and only lines with N$_{\rm HI}$/cm$^{-2}=10^{13-14}$  and
  for which the relative error on the $b$-parameter is smaller than
  $0.5$ are used for all data shown. {\it Right.} Distribution of the
  volume weighted gas temperature when selecting gas with
  overdensities in the range $\delta=4-40$.}
\label{fig1}
\end{figure*}

\begin{figure*}
\begin{center}
\includegraphics[width=7.425cm,height=7.75cm]{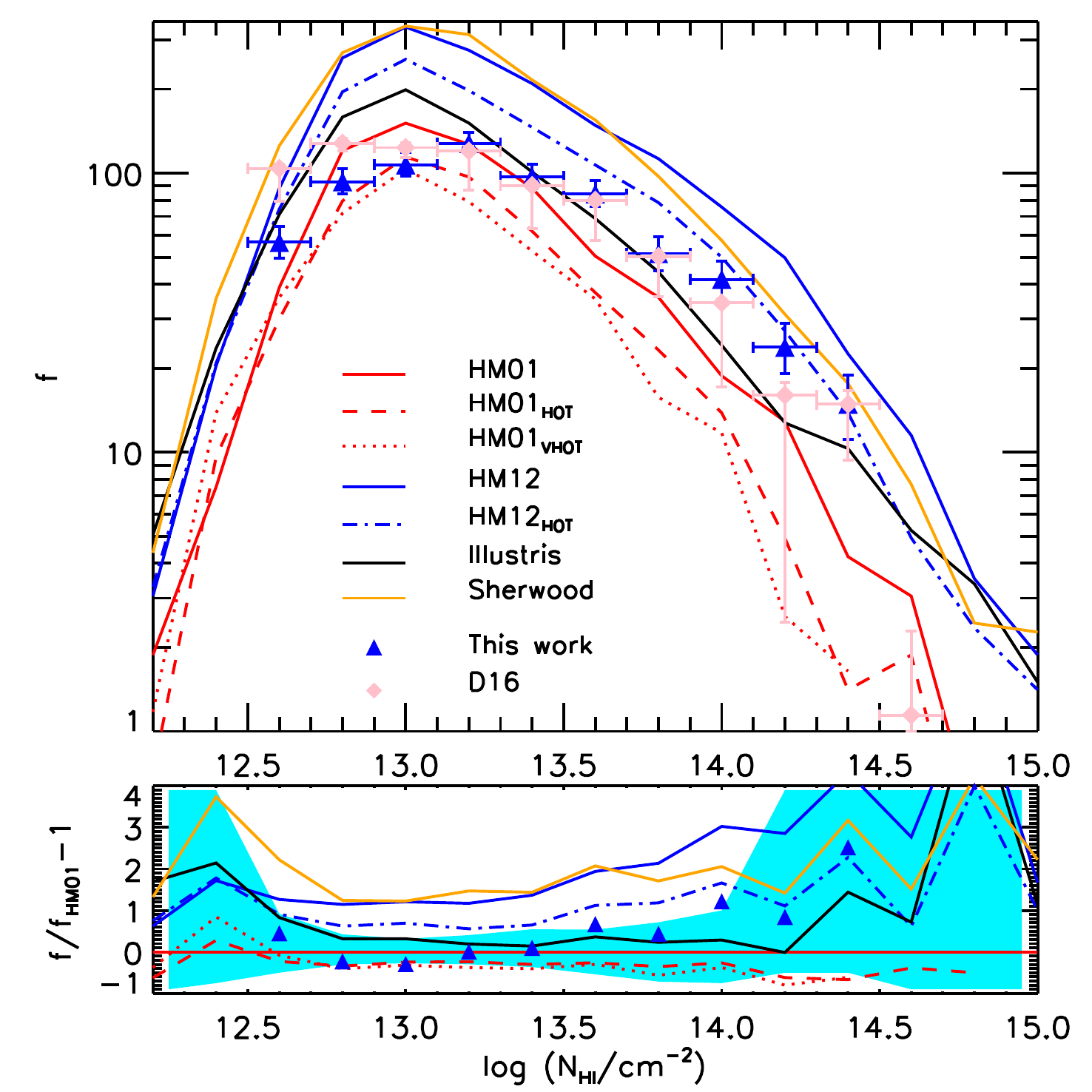}
\includegraphics[width=7.425cm,height=7.75cm]{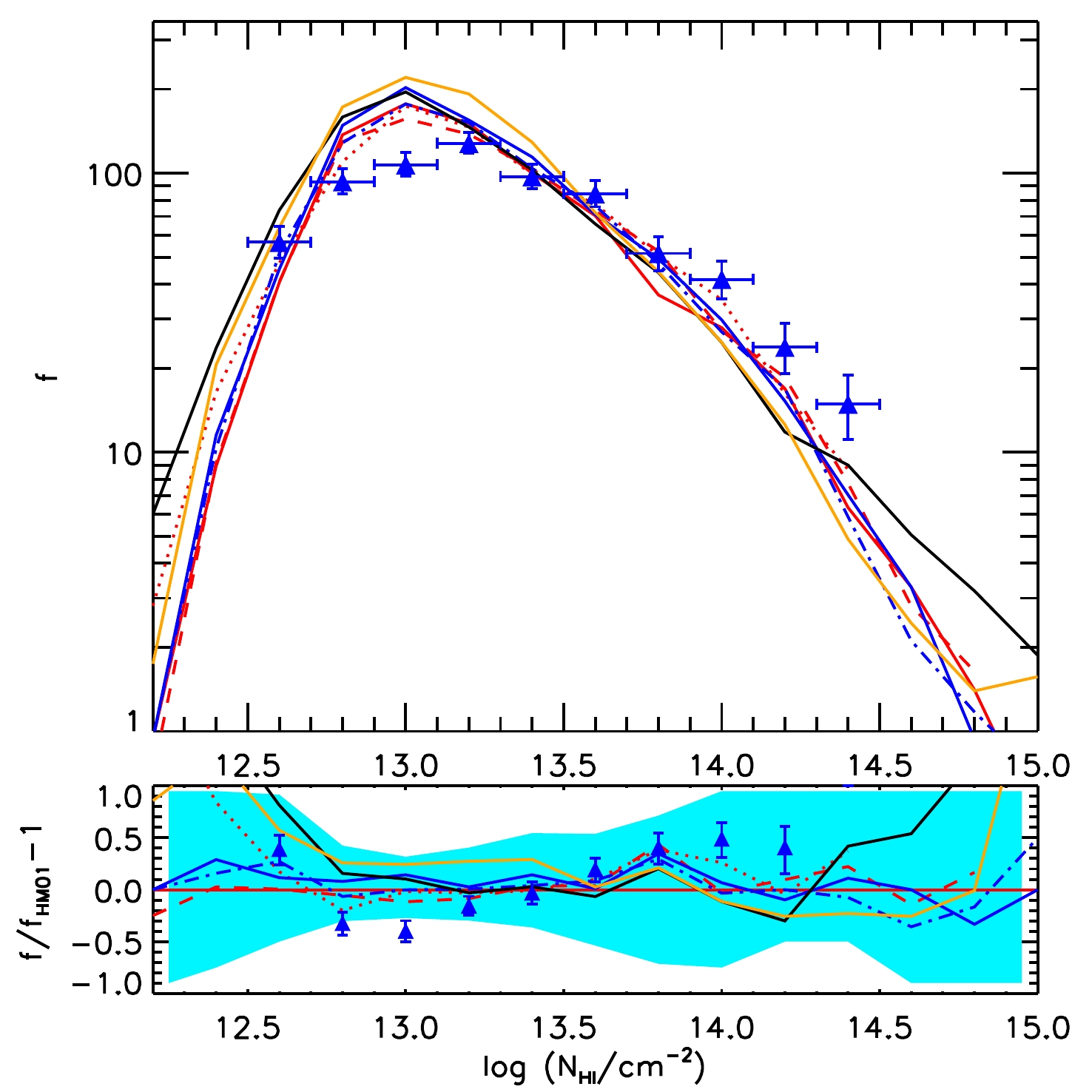}
\end{center}
\vspace{-0.4cm}
\caption{{\it Left.} The corresponding CDDF (log. scale) for the data
  described in Fig. 1 with the addition of the HM12 model (solid blue
  curve).  No scaling has been applied to the mean transmission of the
  simulated spectra. {\it Right.}  The effect of scaling the optical
  depths (and hence \HI photo-ionisation rate) to fit the CDDF in the
  range $\log(N_{\mathrm{\ion{H}{i}}}$/cm$^{-2}$)=13-14.  Data are affected
  by incompleteness at $\log(N_{\mathrm{\ion{H}{i}}}$/cm$^{-2}$) $\le 13$.}
\label{fig_n}
\end{figure*}


In Fig.~\ref{fig1} (left) we show the main result of the present work:
the line width distributions for the simulations and COS data.  It is
clear that HM01 and Illustris do not provide a good fit to the data.
The most problematic ranges are at $b=25-45$ $\rm km\,s^{-1}$, where
HM01 and Illustris underpredict the number of lines by roughly a
factor of two, and below 20 $\rm km\,s^{-1}$ where these models are a
factor of four higher than the data.  The Sherwood simulation is in
better agreement with the data, although it still slightly
overpredicts (underpredicts) the number of lines at $<20 \rm
km\,s^{-1}$ ($b=40$--$60\rm\,km\,s^{-1}$).  We should caution here,
however, that the distribution at $b<20$ $\rm km\,s^{-1}$ is not fully
converged with mass resolution for the HM and Sherwood simulations,
and will slightly underpredict the incidence of narrow lines.
However, this regime is numerically converged for Illustris.  The
median $b-$values are 28.0, 34.5, 36.5 $\rm km\,s^{-1}$ for HM01,
HM01$_{\rm hot}$, HM01$_{\rm vhot}$, respectively; 32.9 $\rm
km\,s^{-1}$ for HM12$_{\rm hot}$ and 28.3 and 33.6 $\rm km\,s^{-1}$
for Illustris and Sherwood, respectively, while the COS data have a
median of 36.2 $\rm km\,s^{-1}$.

Only the HM01$_{\rm hot}$ and HM12$_{\rm hot}$ simulations, which have
been obtained by multiplying the \HeII photo-heating rates by a factor
of three, are in good agreement with the data.  Here HM01$_{\rm hot}$
is around $17\,000\rm\, K$ ($4000\rm\, K$) hotter than HM01 at $z=0.1$
for overdensities $\delta=4-40$ ($\delta=0$). For the corresponding
HM12 simulation the change in temperature is similar.  HM01$_{\rm
  vhot}$, in which the \HeII photo-heating rate has been increased by
a factor of five, is instead too hot and underpredicts the number of
narrow lines with $b<25\rm\,km\,s^{-1}$.

Interestingly, the Illustris simulation is remarkably close to the
HM01 model, despite the considerable differences in the sub-grid
physics used in these simulations.  In the right panel of
Fig.~\ref{fig1} we show the probability distribution function (PDF) of
the gas temperature for overdensities $\delta=4-40$ -- this selects
systems in the column density range considered in this work
\citep[cf.][]{schaye01}. The HM01 model is too cold to produce broad
lines; the HM01$_{\rm vhot}$ model instead has a PDF peaking at
$10^{4.35}\rm\, K$ and in general more gas in the range
$10^{4.25-5}\rm\,K$ due to enhanced photo-heating.  The Sherwood run
has temperatures closer to that of HM01$_{\rm vhot}$ and HM01$_{\rm
  hot}$ runs, although also exhibits a peak at $10^{5}\rm\,K$ arising
from galactic feedback.  In contrast, the Illustris simulation shows
much higher temperatures, with a PDF that peaks at $10^{6.5}\rm\,K$;
this hot gas is too collisionally ionised to produce \lya absorption,
resulting in a similar $b$-parameter distribution to HM01. This is due
to the very energetic AGN bubble injections in Illustris which drive
strong shocks that travel into the IGM and fill most of the volume at $z\sim 0.1$.

Finally, we have also analysed further simulations not presented in
Fig.~\ref{fig1} with a wider range of feedback implementations. A
Sherwood run with only stellar feedback results in an increase by
roughly $3 \rm km\,s^{-1}$ in the peak of the $b-$parameter
distribution with respect to a {\tt QUICKLYA} model, while the
implementation of AGN feedback (orange solid line in Fig.~\ref{fig1})
increases this value further by another $2 \rm\,km\,s^{-1}$.
Similarly, an increase of $4\rm km\,s^{-1}$ in the peak of the
distribution was found when comparing a kinetic wind implementation
with $480 \rm km\,s^{-1}$ winds with the HM01 run.  This demonstrates
the impact of stellar and AGN feedback on the IGM temperature
distribution is strong, and suggests the \lya forest $b$-parameter
distribution is a useful diagnostic of galactic feedback in the low
redshift Universe.


In the left panel of Fig.~\ref{fig_n} we also compare the CDDF, $f=d^2
N/ d \log N_{\rm HI} dz$, of the simulations to the COS data.  The
HM01 and the Illustris simulations -- the latter uses the
\citet{faucher09} UVB model -- are in good agreement with the data in
the range $\log N_{\mathrm{\ion{H}{i}}}$/cm$^{-2}$=13-14, while the
Sherwood and HM12 runs overpredict the number of absorption systems by
a factor of $\sim 2$. The HM12$_{\rm hot}$ model results in better
agreement (since the neutral hydrogen fraction scales approximately
$T^{-0.7}$ through the recombination coefficient) but still lies
significantly above the data.  In this comparison, there is no
rescaling of optical depths and these simulations have values of
$\Gamma$ and $\overline{F}$ as summarised in Table~\ref{tabsims}
(cf. $\overline{F}=0.983$ from D16).

In the right panel we show what happens when we require the
simulations to fit the CDDF in the range we consider most robust,
$\log(N_{\mathrm{\ion{H}{i}}}$/cm$^{-2})=13-14$, by rescaling the
optical depths in the mock spectra.  The values of the mean
transmitted flux, $\overline{F}_{\rm f}$, and photo-ionisation rate,
$\Gamma_{\rm f}$, inferred are listed in Table~\ref{tabsims}.  Overall,
we find that the mean transmitted fluxes are in the range
$\overline{F}_{\rm f}=[0.977-0.982]$, in good agreement with the D16
value (having verified that matching the D16 mean transmitted flux
or the CDDF at these column densities is roughly equivalent), and the
inferred photo-ionisation rates are in the range $\Gamma_{\rm f} =
[0.05-0.1] \times 10^{-12} \rm\,s^{-1}$ (these values must be compared
to the original UVB values $\Gamma$ used as an input for the simulations see Table 1).  The latter are a factor
$1.5$--$3$ higher than predicted by the widely used HM12 UVB model and
are in very good agreement with recent results \citep{shull15,khaire2,cristiani16,gaikwada,gaikwad_aa}.
Note also  that the effects of feedback appear to be more prominent for
absorbers with column densities $\log(N_{\mathrm{\ion{H}{i}}}$/cm$^{-2})>14.5$.

\section{Conclusions}
We have used hydrodynamic simulations to explore several properties of
the \lya forest at $z=0.1$: the $b$-parameter distribution, CDDF and
mean transmitted flux.  The simulations probe a wide range of
different UVBs, feedback and star formation implementations, box sizes
and resolutions.  We find that several simulations fail in reproducing
the line width distribution, under-predicting the number of lines with
$b$-parameter values 25-45 $\rm km\,s^{-1}$ by a factor of two when
compared to the observational data.  This is either because the gas is
too cold or, in models with vigorous AGN feedback, collisionally
ionised. This tension is partly alleviated when considering
alternative feedback models (less aggressive AGN feedback and galactic winds) used in the Sherwood run; it only
disappears in an {\it ad-hoc} model with enhanced photo-heating,
resulting in a median temperature $10^5\rm\,K$ for the IGM with
overdensities $\delta=[4-40]$. 

The CDDF and mean flux are furthermore reproduced only if the
photo-ionisation rate is higher than predicted by the HM12 model by at
least a factor 1.5.  The discrepancy between the values of the
photo-ionisation rate required to match the COS data with those
predicted by the HM12 model is around a factor of 2, rather than the
factor 5 suggested by \citep{kollmeier14}.  This is largely due to the
presence of hot(ter) gas in our simulations.  Overall, we conclude that
comparison of models to the observed $b$-parameter distributions
provides a valuable diagnostic of feedback in the low redshift IGM,
and may help pinpoint any missing physical ingredients in current
hydrodynamic simulations in the form of additional or different
thermal feedback or turbulence \citep[e.g][]{iapichino}.

\section*{Acknowledgments.}
MV is supported by INFN/PD51 Indark,  and (with
TSK) ERC Grant 257670-cosmoIGM.  JSB is supported by a Royal Society
URF. MGH and EP acknowledge support from the FP7 ERC Grant
Emergence-320596 and the Kavli Foundation.  Simulations were performed
at the University of Cambridge with Darwin-HPCS and COSMOS, operated
on behalf of the STFC DiRAC facility (funded by BIS National
E-infrastructure capital grant ST/J005673/1 and STFC grants
ST/H008586/1, ST/K00333X/1), and on the Curie supercomputer at TGCC
through the 8th PRACE call. BPW is supported by NASA grants HST-AR-12842,
HST-AR-13893  from STSCI  operated by AURA under contract NAS5-26555 and AST-1108913 by NSF.

\bibliographystyle{mn2e}

\begin{thebibliography}{}


\bibitem[\protect\citeauthoryear{{Bolton}, {Puchwein}, {Sijacki}, {Haehnelt},
  {Kim}, {Meiksin}, {Regan} \& {Viel}}{{Bolton} et~al.}{2017}]{bolton17}
{Bolton} J.~S.,  {Puchwein} E.,  {Sijacki} D.,  {Haehnelt} M.~G.,  {Kim} T.-S.,
   {Meiksin} A.,  {Regan} J.~A.,    {Viel} M.,  2017, \mnras, 464, 897

\bibitem[\protect\citeauthoryear{{Bolton}, {Viel}, {Kim}, {Haehnelt} \&
  {Carswell}}{{Bolton} et~al.}{2008}]{bolton08pdf}
{Bolton} J.~S.,  {Viel} M.,  {Kim} T.-S.,  {Haehnelt} M.~G.,    {Carswell}
  R.~F.,  2008, \mnras, 386, 1131

\bibitem[\protect\citeauthoryear{{Carswell} \& {Webb}}{{Carswell} \&
  {Webb}}{2014}]{carswell14}
{Carswell} R.~F.,  {Webb} J.~K.,  2014, VPFIT, Astrophysics Source Code
  Library, record ascl:1408.015

\bibitem[\protect\citeauthoryear{{Cristiani}, {Serrano}, {Fontanot}, {Vanzella}
  \& {Monaco}}{{Cristiani} et~al.}{2016}]{cristiani16}
{Cristiani} S.,  {Serrano} L.~M.,  {Fontanot} F.,  {Vanzella} E.,    {Monaco}
  P.,  2016, \mnras, 462, 2478

\bibitem[\protect\citeauthoryear{{Danforth}, {Keeney}, {Tilton}, {Shull},
  {Stocke}, {Stevans}, {Pieri}, {Savage}, {France}, {Syphers}, {Smith},
  {Green}, {Froning}, {Penton} \& {Osterman}}{{Danforth}
  et~al.}{2016}]{danforth16}
{Danforth} C.~W.,  {Keeney} B.~A.,  {Tilton} E.~M.,  {Shull} J.~M.,  {Stocke}
  J.~T.,  {Stevans} M.,  {Pieri} M.~M.,  {Savage} B.~D.,  {France} K.,
  {Syphers} D.,  {Smith} B.~D.,  {Green} J.~C.,  {Froning} C.,  {Penton} S.~V.,
     {Osterman} S.~N.,  2016, \apj, 817, 111

\bibitem[\protect\citeauthoryear{{Dav{\'e}}, {Oppenheimer}, {Katz}, {Kollmeier}
  \& {Weinberg}}{{Dav{\'e}} et~al.}{2010}]{dave10}
{Dav{\'e}} R.,  {Oppenheimer} B.~D.,  {Katz} N.,  {Kollmeier} J.~A.,
  {Weinberg} D.~H.,  2010, \mnras, 408, 2051

\bibitem[\protect\citeauthoryear{{Faucher-Gigu{\`e}re}, {Lidz}, {Zaldarriaga}
  \& {Hernquist}}{{Faucher-Gigu{\`e}re} et~al.}{2009}]{faucher09}
{Faucher-Gigu{\`e}re} C.-A.,  {Lidz} A.,  {Zaldarriaga} M.,    {Hernquist} L.,
  2009, \apj, 703, 1416

\bibitem[\protect\citeauthoryear{{Ford}, {Oppenheimer}, {Dav{\'e}}, {Katz},
  {Kollmeier} \& {Weinberg}}{{Ford} et~al.}{2013}]{ford13}
{Ford} A.~B.,  {Oppenheimer} B.~D.,  {Dav{\'e}} R.,  {Katz} N.,  {Kollmeier}
  J.~A.,    {Weinberg} D.~H.,  2013, \mnras, 432, 89

\bibitem[\protect\citeauthoryear{{Gaikwad}, {Khaire}, {Choudhury} \&
  {Srianand}}{{Gaikwad} et~al.}{2016a}]{gaikwada}
{Gaikwad} P.,  {Khaire} V.,  {Choudhury} T.~R.,    {Srianand} R.,  2016, ArXiv
  e-prints: 1605.02738

\bibitem[\protect\citeauthoryear{{Gaikwad}, {Srianand}, {Choudhury} \&
  {Khaire}}{{Gaikwad} et~al.}{2016b}]{gaikwad_aa}
{Gaikwad} P.,  {Srianand} R.,  {Choudhury} T.~R.,    {Khaire} V.,  2016, ArXiv
  e-prints: 1610.06572

\bibitem[\protect\citeauthoryear{{Gurvich}, {Burkhart} \& {Bird}}{{Gurvich}
  et~al.}{2016}]{bird16}
{Gurvich} A.,  {Burkhart} B.,    {Bird} S.,  2016, ArXiv e-prints: 1608.03293

\bibitem[\protect\citeauthoryear{{Haardt} \& {Madau}}{{Haardt} \&
  {Madau}}{2001}]{hm01}
{Haardt} F.,  {Madau} P.,  2001, astro-ph/0106018

\bibitem[\protect\citeauthoryear{{Haardt} \& {Madau}}{{Haardt} \&
  {Madau}}{2012}]{hm12}
{Haardt} F.,  {Madau} P.,  2012, \apj, 746, 125

\bibitem[\protect\citeauthoryear{{Hinshaw}, {Larson}, {Komatsu}
  et~al.,}{{Hinshaw} et~al.}{2013}]{wmap9yr}
{Hinshaw} G.,  {Larson} D.,  {Komatsu} E.,    et~al., 2013, \apjs, 208, 19

\bibitem[\protect\citeauthoryear{{Iapichino}, {Viel} \& {Borgani}}{{Iapichino}
  et~al.}{2013}]{iapichino}
{Iapichino} L.,  {Viel} M.,    {Borgani} S.,  2013, \mnras, 432, 2529

\bibitem[\protect\citeauthoryear{{Janknecht}, {Reimers}, {Lopez} \&
  {Tytler}}{{Janknecht} et~al.}{2006}]{janknecht06}
{Janknecht} E.,  {Reimers} D.,  {Lopez} S.,    {Tytler} D.,  2006, \aap, 458,
  427

\bibitem[\protect\citeauthoryear{{Khaire} \& {Srianand}}{{Khaire} \&
  {Srianand}}{2015}]{khaire}
{Khaire} V.,  {Srianand} R.,  2015, \mnras, 451, L30

\bibitem[\protect\citeauthoryear{{Khaire}, {Srianand}, {Choudhury} \&
  {Gaikwad}}{{Khaire} et~al.}{2016}]{khaire2}
{Khaire} V.,  {Srianand} R.,  {Choudhury} T.~R.,    {Gaikwad} P.,  2016,
  \mnras, 457, 4051

\bibitem[\protect\citeauthoryear{{Kim}, {Carswell}, {Mongardi}, {Partl},
  {M\"ucket}, {Barai} \& {Cristiani}}{{Kim} et~al.}{2016}]{kim16}
{Kim} T.-S.,  {Carswell} R.~F.,  {Mongardi} C.,  {Partl} A.~M.,  {M\"ucket}
  J.~P.,  {Barai} P.,    {Cristiani} S.,  2016, MNRAS, 457, 2005

\bibitem[\protect\citeauthoryear{{Kim}, {Partl}, {Carswell} \&
  {M{\"u}ller}}{{Kim} et~al.}{2013}]{kim13}
{Kim} T.-S.,  {Partl} A.~M.,  {Carswell} R.~F.,    {M{\"u}ller} V.,  2013,
  \aap, 552, A77

\bibitem[\protect\citeauthoryear{{Kirkman}, {Tytler}, {Lubin} \&
  {Charlton}}{{Kirkman} et~al.}{2007}]{kirkman07}
{Kirkman} D.,  {Tytler} D.,  {Lubin} D.,    {Charlton} J.,  2007, \mnras, 376,
  1227

\bibitem[\protect\citeauthoryear{{Kollmeier}, {Weinberg}, {Oppenheimer},
  {Haardt}, {Katz}, {Dav{\'e}}, {Fardal}, {Madau}, {Danforth}, {Ford},
  {Peeples} \& {McEwen}}{{Kollmeier} et~al.}{2014}]{kollmeier14}
{Kollmeier} J.~A.,  {Weinberg} D.~H.,  {Oppenheimer} B.~D.,  {Haardt} F.,
  {Katz} N.,  {Dav{\'e}} R.,  {Fardal} M.,  {Madau} P.,  {Danforth} C.,  {Ford}
  A.~B.,  {Peeples} M.~S.,    {McEwen} J.,  2014, \apjl, 789, L32

\bibitem[\protect\citeauthoryear{{Kriss}}{{Kriss}}{2011}]{kriss11}
{Kriss} G.~A.,  2011, COS Instrument Science Report 2011-01(v1), 17 pages, p.~1

\bibitem[\protect\citeauthoryear{{Nelson}, {Pillepich}, {Genel},
  {Vogelsberger}, {Springel}, {Torrey}, {Rodriguez-Gomez}, {Sijacki}, {Snyder},
  {Griffen}, {Marinacci}, {Blecha}, {Sales}, {Xu} \& {Hernquist}}{{Nelson}
  et~al.}{2015}]{nelson}
{Nelson} D.,  {Pillepich} A.,  {Genel} S.,  {Vogelsberger} M.,  {Springel} V.,
  {Torrey} P.,  {Rodriguez-Gomez} V.,  {Sijacki} D.,  {Snyder} G.~F.,
  {Griffen} B.,  {Marinacci} F.,  {Blecha} L.,  {Sales} L.,  {Xu} D.,
  {Hernquist} L.,  2015, Astronomy and Computing, 13, 12

\bibitem[\protect\citeauthoryear{{Pachat}, {Narayanan}, {Muzahid}, {Khaire},
  {Srianand}, {Wakker} \& {Savage}}{{Pachat} et~al.}{2016}]{pachat}
{Pachat} S.,  {Narayanan} A.,  {Muzahid} S.,  {Khaire} V.,  {Srianand} R.,
  {Wakker} B.~P.,    {Savage} B.~D.,  2016, \mnras, 458, 733

\bibitem[\protect\citeauthoryear{{Planck Collaboration}, {Ade}, {Aghanim},
  {Armitage-Caplan}, {Arnaud}, {Ashdown}, {Atrio-Barandela}, {Aumont},
  {Baccigalupi}, {Banday} \& et al.}{{Planck Collaboration}
  et~al.}{2014}]{Planck2014}
{Planck Collaboration} {Ade} P.~A.~R.,  {Aghanim} N.,  {Armitage-Caplan} C.,
  {Arnaud} M.,  {Ashdown} M.,  {Atrio-Barandela} F.,  {Aumont} J.,
  {Baccigalupi} C.,  {Banday} A.~J.,    et al. 2014, \aap, 571, A16

\bibitem[\protect\citeauthoryear{{Puchwein} \& {Springel}}{{Puchwein} \&
  {Springel}}{2013}]{puchwein13}
{Puchwein} E.,  {Springel} V.,  2013, \mnras, 428, 2966

\bibitem[\protect\citeauthoryear{{Rahmati}, {Schaye}, {Crain}, {Oppenheimer},
  {Schaller} \& {Theuns}}{{Rahmati} et~al.}{2016}]{rahmati16}
{Rahmati} A.,  {Schaye} J.,  {Crain} R.~A.,  {Oppenheimer} B.~D.,  {Schaller}
  M.,    {Theuns} T.,  2016, \mnras, 459, 310

\bibitem[\protect\citeauthoryear{{Savage}, {Kim}, {Wakker}, {Keeney}, {Shull},
  {Stocke} \& {Green}}{{Savage} et~al.}{2014}]{savage14}
{Savage} B.~D.,  {Kim} T.-S.,  {Wakker} B.~P.,  {Keeney} B.,  {Shull} J.~M.,
  {Stocke} J.~T.,    {Green} J.~C.,  2014, \apjs, 212, 8

\bibitem[\protect\citeauthoryear{{Schaye}}{{Schaye}}{2001}]{schaye01}
{Schaye} J.,  2001, \apj, 559, 507

\bibitem[\protect\citeauthoryear{{Shull}, {Danforth} \& {Tilton}}{{Shull}
  et~al.}{2014}]{shull14}
{Shull} J.~M.,  {Danforth} C.~W.,    {Tilton} E.~M.,  2014, \apj, 796, 49

\bibitem[\protect\citeauthoryear{{Shull}, {Moloney}, {Danforth} \&
  {Tilton}}{{Shull} et~al.}{2015}]{shull15}
{Shull} J.~M.,  {Moloney} J.,  {Danforth} C.~W.,    {Tilton} E.~M.,  2015,
  \apj, 811, 3

\bibitem[\protect\citeauthoryear{{Sijacki}, {Springel}, {Di Matteo} \&
  {Hernquist}}{{Sijacki} et~al.}{2007}]{sijacki07}
{Sijacki} D.,  {Springel} V.,  {Di Matteo} T.,    {Hernquist} L.,  2007,
  \mnras, 380, 877

\bibitem[\protect\citeauthoryear{{Springel}}{{Springel}}{2005}]{springel}
{Springel} V.,  2005, \mnras, 364, 1105

\bibitem[\protect\citeauthoryear{{Springel}}{{Springel}}{2010}]{springel10}
{Springel} V.,  2010, \mnras, 401, 791

\bibitem[\protect\citeauthoryear{{Springel} \& {Hernquist}}{{Springel} \&
  {Hernquist}}{2003}]{springel2003}
{Springel} V.,  {Hernquist} L.,  2003, \mnras, 339, 289

\bibitem[\protect\citeauthoryear{{Tepper-Garc{\'{\i}}a}, {Richter}, {Schaye},
  {Booth}, {Dalla Vecchia} \& {Theuns}}{{Tepper-Garc{\'{\i}}a}
  et~al.}{2012}]{tepper}
{Tepper-Garc{\'{\i}}a} T.,  {Richter} P.,  {Schaye} J.,  {Booth} C.~M.,  {Dalla
  Vecchia} C.,    {Theuns} T.,  2012, \mnras, 425, 1640

\bibitem[\protect\citeauthoryear{{Tornatore}, {Borgani}, {Viel} \&
  {Springel}}{{Tornatore} et~al.}{2010}]{torna10}
{Tornatore} L.,  {Borgani} S.,  {Viel} M.,    {Springel} V.,  2010, \mnras,
  402, 1911

\bibitem[\protect\citeauthoryear{{Viel}, {Haehnelt} \& {Springel}}{{Viel}
  et~al.}{2004}]{vhs}
{Viel} M.,  {Haehnelt} M.~G.,    {Springel} V.,  2004, \mnras, 354, 684

\bibitem[\protect\citeauthoryear{{Villaescusa-Navarro}, {Planelles}, {Borgani},
  {Viel}, {Rasia}, {Murante}, {Dolag}, {Steinborn}, {Biffi}, {Beck} \&
  {Ragone-Figueroa}}{{Villaescusa-Navarro} et~al.}{2016}]{villa16}
{Villaescusa-Navarro} F.,  {Planelles} S.,  {Borgani} S.,  {Viel} M.,  {Rasia}
  E.,  {Murante} G.,  {Dolag} K.,  {Steinborn} L.~K.,  {Biffi} V.,  {Beck}
  A.~M.,    {Ragone-Figueroa} C.,  2016, \mnras, 456, 3553

\bibitem[\protect\citeauthoryear{{Vogelsberger}, {Genel}, {Sijacki}, {Torrey},
  {Springel} \& {Hernquist}}{{Vogelsberger} et~al.}{2013}]{vogelsberger2013}
{Vogelsberger} M.,  {Genel} S.,  {Sijacki} D.,  {Torrey} P.,  {Springel} V.,
  {Hernquist} L.,  2013, \mnras, 436, 3031

\bibitem[\protect\citeauthoryear{{Vogelsberger}, {Genel}, {Springel}, {Torrey},
  {Sijacki}, {Xu}, {Snyder}, {Bird}, {Nelson} \& {Hernquist}}{{Vogelsberger}
  et~al.}{2014}]{illustris}
{Vogelsberger} M.,  {Genel} S.,  {Springel} V.,  {Torrey} P.,  {Sijacki} D.,
  {Xu} D.,  {Snyder} G.,  {Bird} S.,  {Nelson} D.,    {Hernquist} L.,  2014,
  \nat, 509, 177

\bibitem[\protect\citeauthoryear{{Wakker}, {Hernandez}, {French}, {Kim},
  {Oppenheimer} \& {Savage}}{{Wakker} et~al.}{2015}]{wakker15}
{Wakker} B.~P.,  {Hernandez} A.~K.,  {French} D.~M.,  {Kim} T.-S.,
  {Oppenheimer} B.~D.,    {Savage} B.~D.,  2015, \apj, 814, 40

\bibitem[\protect\citeauthoryear{{Werk}, {Prochaska}, {Cantalupo}, {Fox},
  {Oppenheimer}, {Tumlinson}, {Tripp}, {Lehner} \& {McQuinn}}{{Werk}
  et~al.}{2016}]{werk16}
{Werk} J.~K.,  {Prochaska} J.~X.,  {Cantalupo} S.,  {Fox} A.~J.,  {Oppenheimer}
  B.,  {Tumlinson} J.,  {Tripp} T.~M.,  {Lehner} N.,    {McQuinn} M.,  2016,
  \apj, 833, 54

\bibitem[\protect\citeauthoryear{{Weymann}, {Jannuzi}, {Lu}, {Bahcall},
  {Bergeron}, {Boksenberg}, {Hartig}, {Kirhakos}, {Sargent}, {Savage},
  {Schneider}, {Turnshek} \& {Wolfe}}{{Weymann} et~al.}{1998}]{weymann98}
{Weymann} R.~J.,  {Jannuzi} B.~T.,  {Lu} L.,  {Bahcall} J.~N.,  {Bergeron} J.,
  {Boksenberg} A.,  {Hartig} G.~F.,  {Kirhakos} S.,  {Sargent} W.~L.~W.,
  {Savage} B.~D.,  {Schneider} D.~P.,  {Turnshek} D.~A.,    {Wolfe} A.~M.,
  1998, \apj, 506, 1

\end{thebibliography}

\newcommand{\noopsort}[1]{}
{}
 
\end{document}